\documentclass[nopreprintline,12pt]{elsarticle}

\usepackage{amssymb}
\usepackage{amsmath}

\usepackage{lineno}

\usepackage{booktabs}
\usepackage{multirow}

\usepackage{xcolor}
\usepackage{url}
\usepackage{breakurl}
\usepackage[breaklinks, colorlinks, pdfstartview={XYZ null null null}]{hyperref}

\AtBeginDocument{
	\hypersetup{
		linkcolor=.,
		citecolor={blue!80!black},
		urlcolor={black}
	}
}

\journal{Mathematical Methods in the Applied Sciences}

\graphicspath{{Figures/}}

\begin{document}

\begin{frontmatter}


\noindent
This is the peer reviewed version of the following article:
\\\\
Lloret-Climent M, Nescolarde-Selva J-A, Alonso-Stenberg K, Montoyo A, Gutiérrez Y. Applying Smarta to the analysis of tourist networks. Math Meth Appl Sci. 2022; 45(7): 3921–3932. doi:10.1002/mma.8023,
\\\\
which has been published in final form at \url{https://doi.org/10.1002/mma.8023}.
\\\\
This article may be used for non-commercial purposes in accordance with Wiley Terms and Conditions for Use of Self-Archived Versions. This article may not be enhanced, enriched or otherwise transformed into a derivative work, without express permission from Wiley or by statutory rights under applicable legislation. Copyright notices must not be removed, obscured or modified. The article must be linked to Wiley’s version of record on Wiley Online Library and any embedding, framing or otherwise making available the article or pages thereof by third parties from platforms, services and websites other than Wiley Online Library must be prohibited.
\title{Applying Smarta to the analysis of tourist networks\tnoteref{support}}
\tnotetext[support]{This work was supported by the Ministry of Economy and Business (Government of Spain), under the project RTI2018-094653-B-C22; and the University of Alicante, under the project UAPOSTCOVID19-11.}

\author[label1]{Miguel Lloret-Climent}
\ead{miguel.lloret@ua.es}
\author{Josué Antonio Nescolarde-Selva\corref{cor1}\fnref{label1}}
\ead{josue.selva@ua.es}
\author[label1]{Kristian Alonso-Stenberg}
\ead{kristian.alonso@ua.es}
\author[label2]{Andrés Montoyo-Guijarro}
\ead{montoyo@ua.es}
\author[label2]{Yoan Gutiérrez-Vázquez}
\ead{ygutierrez@ua.es}
\address[label1]{Department of Applied Mathematics, University of Alicante, Alicante 03690, Spain}
\address[label2]{Department of Software and Computing Systems, University of Alicante, Alicante 03690, Spain}
\cortext[cor1]{Corresponding author at: Department of Applied Mathematics, University of Alicante, Alicante 03690, Spain.}
\tnotetext[]{Declarations of interest: none.}

\begin{abstract}
	
The framework of the present study was the destination life cycle model, a classical model that describes the development of tourist destinations. We examined mass tourism in Benidorm based on tourist accommodation supply and demand statistics over the January 2016 - October 2018 period, provided by Spain’s National Institute for Statistics. The objective was to analyze the life cycle and competitiveness of Benidorm’s tourism system, interpret whether the tourism product was sustainable, and at what stage in the cycle Benidorm is currently in. To do this, we used Smarta software, which, based on network analysis, enables to interpret the system’s virtuous cycles and analyze causality by observing relationship patterns in the system's attractors, thus complementing typical processing based on causal maps and the study of social networks. The results obtained by this application (which has been developed by our research team), show 6 sets of attractors that mark the trends of the tourist system. Finally, the analysis of the significant variables of these attractors have helped us to justify that the tourist system of Benidorm is in the rejuvenation phase.

\end{abstract}



	
\end{frontmatter}


\section{Introduction}
\label{sec:introduction}

Many complex systems, such as tourism systems, can be described as networks with interacting elements and using terms based on graph theory. Network analysis is a mathematical tool that allows characterising the structure of complex networks, connecting their structural and functional features. Authors such as \cite{albert2002,boccaletti2006,newman2003} analyzed topological aspects of many social and natural networks. Several studies in the field of tourism have applied network analysis, such as \cite{baggio2010,delchiappa2015,casanueva2016,lozano2018,lloret2018b}. These latter works mainly studied tourism destination networks. The main goal of the present study was to analyze the relationships between a tourist destination’s multiple stakeholders, examining the density and clusters of network relationships and identifying the interesting parts that play an essential role.

Benidorm is a long-established sun and beach destination on Spain’s Mediterranean coast. The city has around 80,000 inhabitants, but it hosts almost two million visitors and over ten thousand overnight stays a year, leading the city to be overcrowded, especially in the summer months. The city is built in a peculiar way and is very different from other European cities: it has the highest density of skyscrapers per inhabitant in the world, and the second highest density per square meter after New York. The population increase is mainly due to the large number of tourists arriving in the city in summer, supplemented by senior tourism in winter.

Tourism and construction are two key sectors of Spain’s economy. Together they have driven the economy since the 1960s. The result so far has been a highly questionable tourist pattern from an environmental perspective. It is a fact that new infrastructures and facilities related to communications (roads, railways, airports, etc.) have been created; and tourism has often been the perfect excuse to justify low sustainable growth. Tourism has become a major predator of coastal area space of high environmental value, causing critical impacts on the environment. A characteristic model regarding the evolution of tourist destinations is the Butler model \cite{butler1980}, which was preceded in the field of business economy by the work of Vernon \cite{vernon1966}, who described how innovations and technological progress affected evolutionary economic cycles. According to this theory, massification and carrying capacity influence the decline of a tourist destination. In the field of tourism, the concept of carrying capacity is interpreted as a strategy to reduce the impact of visitors. Yun and Zhang \cite{yun2017} examined residents' perceptions of impacts and tourism development. The case of Benidorm was studied by Sánchez-Galiano et al. \cite{sanchez2017}, who put forward a population density index to assess the destination’s urban sustainability. The main symptom of decline is the drop in the number of tourists, but other symptoms also include a decrease in: benefits, environmental quality, services and facilities. Soares et al. \cite{soares2012} advanced that Butler’s model did not help to predict a destination’s decline and the authors proposed internal and external factors that could in fact achieve this prediction goal. They suggested analyzing supply and demand variables to understand the point reached by the destination in the evolutionary phase, which was the aim of the present study.

\section{Network analysis: causal maps and Smarta software}
\label{sec:network_analysis}

Causal maps allow visualising and analyzing complex structures of causes and effects by graphically representing variables and relationships \cite{weick1979,gilmore1991,weber2001}. The network’s vertices are connected by unidirectional arrows illustrating causality and are constructed from individual interviews and/or group meetings \cite{ackermann2001}. These causal maps have also been used in the domain of tourism by authors such as: \cite{woodside2009}, who used them in the field of golf; \cite{nash2006}, who applied them to rural areas; \cite{aledo2010}, who applied them to residential tourism; or \cite{baggio2010}, who used them to manage the knowledge and structure of destinations. Causal maps are analyzed using computer programs such as Ucinet and NetDraw \cite{borgatti2002,borgatti2014}; they allow us to analyze social networks by examining the interactions between all the actors.

In the present study, Benidorm, as a dynamic entity, can be interpreted as a union of connected elements. The main issue is to understand how the city’s associated network evolves over time and/or to check how internal mechanisms or external processes such as Brexit can produce changes in its structure, changing the types of relationships between the variables. To do this, we replaced the causal maps with Smarta software \cite{lloret2018,lloret2019,alonso2019}, which is based on network analysis, and allows varying the destination’s data period. This way, we could detect variations dynamically; the tool can thus be understood as a tool that diagnoses the situation of the tourist destination cycle. The network nodes consist in supply and demand variables of three of Benidorm’s accommodation sectors, during the period January 2016 - October 2018. The interactions between these variables (number of travellers, overnight stays, number of establishments, occupancy rate, etc.) show us the state of the destination and they can anticipate possible turbulence.
Smarta offers a method to systemically analyze the causes and effects of the major main actors of a mature destination, though with notable differences:

\begin{itemize}
	\item Causal maps are constructed from individual interviews and/or group meetings, while Smarta obtains causal relationships through conditioned probabilities.
	
	\item The relationships between the variables may be unidirectional, bidirectional or unrelated. We are implicitly assuming a circular causality between the variables, which has already been studied in tourism. For example, \cite{koo2017} analyzed the circular causation between direct air links and tourism demand.
	
	\item The relationships obtained from the programme are directed, no weights are associated with the variables, although weights will be introduced in future versions of the programme.
	
	\item The network’s usual topological characteristics, such as centrality, proximity, etc. are not analyzed, but they are supplemented by the analysis of elements associated with chaos theory such as orbits, invariant sets and especially attractors, which better describe the properties of indirect causality.
\end{itemize}

To end this section, we must highlight that the Smarta application has been developed by our research team (Systemic, Cybernetics and Optimization, SCO), and although its development began in 2009, the application has been improved from 2016.

\section{Methodology}
\label{sec:methodology}

The methodology carried out in this study can be divided into 6 steps:
\begin{enumerate}
	\item Obtaining the database of tourism variables.
	\item Calculation of Pearson correlation between variables.
	\item Determination of cause-effect pairs.
	\item Representation of the associated directed graph.
	\item Obtaining and interpretation of the system attractors.
	\item Analysis of the tourism life cycle of the system.
\end{enumerate}

Next, we proceed to summarize each of the steps mentioned (leaving the detailed justification for later sections):

Step 1 consists of obtaining the tourism variables of Benidorm, from the official website of the National Institute of Statistics (INE) \cite{ine_hotels,ine_campsite,ine_dwellings}. In this sense, we selected the database of occupancy surveys (period January 2016 - October 2018) for 3 accommodation categories: hotels, campsites and holiday flats (dwellings).

Step 2 focuses on calculating the Pearson correlation ($r$) between all possible pairs of variables. If a pair of variables has a correlation higher than the threshold 0.935 (or lower than -0.935, in case of negative correlation), then it will pass the test. If the pair of variables does not meet these conditions, we discard it.

In step 3 we determine the cause-effect pairs of the system. After this process, we will know the causal relationship that exists between each pair of variables that passed the correlation test. For example, if we have a pair of correlated variables $(A,B)$, we will know if $A \rightarrow B$ ($A$ causes $B$), if $B \rightarrow A$ ($B$ causes $A$), if $A \leftrightarrow B$ (bidirectional causality), or if there is no causal relationship.

To better visualize the cause-effect pairs obtained, in step 4 we plot a 2D directed graph. In this representation, we symbolize each variable with a circle, and each relationship with an arrow. In addition, if the correlation is positive the arrow will be black, and if the correlation is negative the arrow will be red.

Step 5 consists of identifying the attractor sets of the system. Attractors are sets of variables in the form of loops, which causally ``attract'' all the variables found in their basin of attraction. Thanks to the attractors obtained, we can study the main trends of the tourism system under study.

In step 6 we analyze the tourist life cycle of the system, based on the study of the most significant variables of the attractors obtained. This analysis will allow us to justify in which phase of the tourist cycle Benidorm is.

Finally, thanks to the Smarta causal simulator, we will be able to automate the whole process from steps 2-5.

\section{Case study}
\label{sec:case_study}

In this section, we analyze the trends and attractors of Benidorm's tourism system, by examining various databases published by the National Institute of Statistics (INE) \cite{ine_hotels,ine_campsite,ine_dwellings}, that collected occupancy information for three accommodation categories: hotels, campsites and holiday flats (dwellings). Furthermore, we conduct an interpretation of Benidorm’s tourist life cycle, observing the relationships between the attractors found over the study period. This period covers from January 2016 to October 2018. The justification for this period is twofold:

\begin{enumerate}
	\item There have been significant changes in Benidorm's tourist occupancy indicators during the period January 2016 - October 2018. Specifically, factors such as the Spanish economic recovery and the value for money of this destination have propelled tourism growth by the Spanish population.
	
	\item  We wish to reserve the data corresponding to the period 2019-2021 for a future article, with the aim of comparing trends, attractors and the tourism life cycle of Benidorm in the pre-COVID and post-COVID era.
\end{enumerate}

\subsection{Analysis of the system’s trends and attractors}
\label{subsec:attractors_analysis}

The main objective of the database is to study the behaviour of different supply and demand variables that allow describing the hotel sector, the camping segment and the holiday flat sector. In this sense, the databases are made up of a total of 35 variables grouped into 6 categories, namely: travellers (guests), overnight stays, supply, personnel employed, occupancy rate and average stay. Of the 35 variables, 12 corresponded to the hotel sector, 12 to camping services and 11 to holiday flats. We assigned a code formed by 3 uppercase letters to each variable to simplify the names of the variables in the study. The variables are described below, and all the information is gathered in Table \ref{tab:1}.

\begin{table}[t!]
	\centering
	\caption{Characteristics of the variables of the database under study (Jan. 2016 - Oct. 2018).}
	\label{tab:1}
	{\scriptsize
		\begin{tabular}{c l c}
			\toprule
			\textit{Variable category} &  \multicolumn{1}{c}{\textit{Variable name}} & \textit{Variable code}\\
			\midrule
			\multirow{6}{*}{Guests} & Hotels: Guests–Residents in Spain & HGS\\
			& Hotels: Guests–Residents abroad & HGA\\
			& Campsites: Guests–Residents in Spain & CGS\\
			& Campsites: Guests–Residents abroad & CGA\\
			& Dwellings: Guests–Residents in Spain & DGS\\
			& Dwellings: Guests–Residents abroad & DGA\\
			\midrule
			\multirow{6}{*}{Overnight stays} & Hotels: Overnight stays–Residents in Spain & HOS\\
			& Hotels: Overnight stays–Residents abroad & HOA\\
			& Campsites: Overnight stays–Residents in Spain & COS\\
			& Campsites: Overnight stays–Residents abroad & COA\\
			& Dwellings: Overnight stays–Residents in Spain & DOS\\
			& Dwellings: Overnight stays–Residents abroad & DOA\\
			\midrule
			\multirow{8}{*}{Supply} & Hotels: Estimated number of open establishments  & HNO\\
			& Hotels: Estimated number of rooms & HNR\\
			& Hotels: Estimated number of bed-places & HNB\\
			& Campsites: Estimated number of open establishments & CNO\\
			& Campsites: Number of plots & CNP\\
			& Campsites: Estimated number of bed-places & CNB\\
			& Dwellings: Estimated number of flats & DNF\\
			& Dwellings: Estimated number of bed-places & DNB\\
			\midrule
			\multirow{3}{*}{Employed personnel} & Hotels: Employed personnel  & HEP\\
			& Campsites: Employed personnel & CEP\\
			& Dwellings: Employed personnel & DEP\\
			\midrule
			\multirow{9}{*}{Occupancy rate} & Hotels: Occupancy rate of bed-places & HOB\\
			& Hotels: Occupancy rate of bed-places at weekends & HOW\\
			& Hotels: Occupancy rate of bedrooms & HOR\\
			& Campsites: Occupied plots & COP\\
			& Campsites: Occupancy rate of plots & COR\\
			& Campsites: Occupancy rate of plots on weekend & COW\\
			& Dwellings: Occupancy rate of bed-places & DOB\\
			& Dwellings: Occupancy rate of bed-places at weekends & DOW\\
			& Dwellings: Occupancy rate of flats & DOF\\
			\midrule
			\multirow{3}{*}{Average stay} &  Hotels: Average stay & HAS\\
			& Campsites: Average stay & CAS\\
			& Dwellings: Average stay & DAS\\
			\bottomrule
		\end{tabular}
	}
\end{table}

\begin{figure}[t!]
	\centering
	\includegraphics[width=\columnwidth]{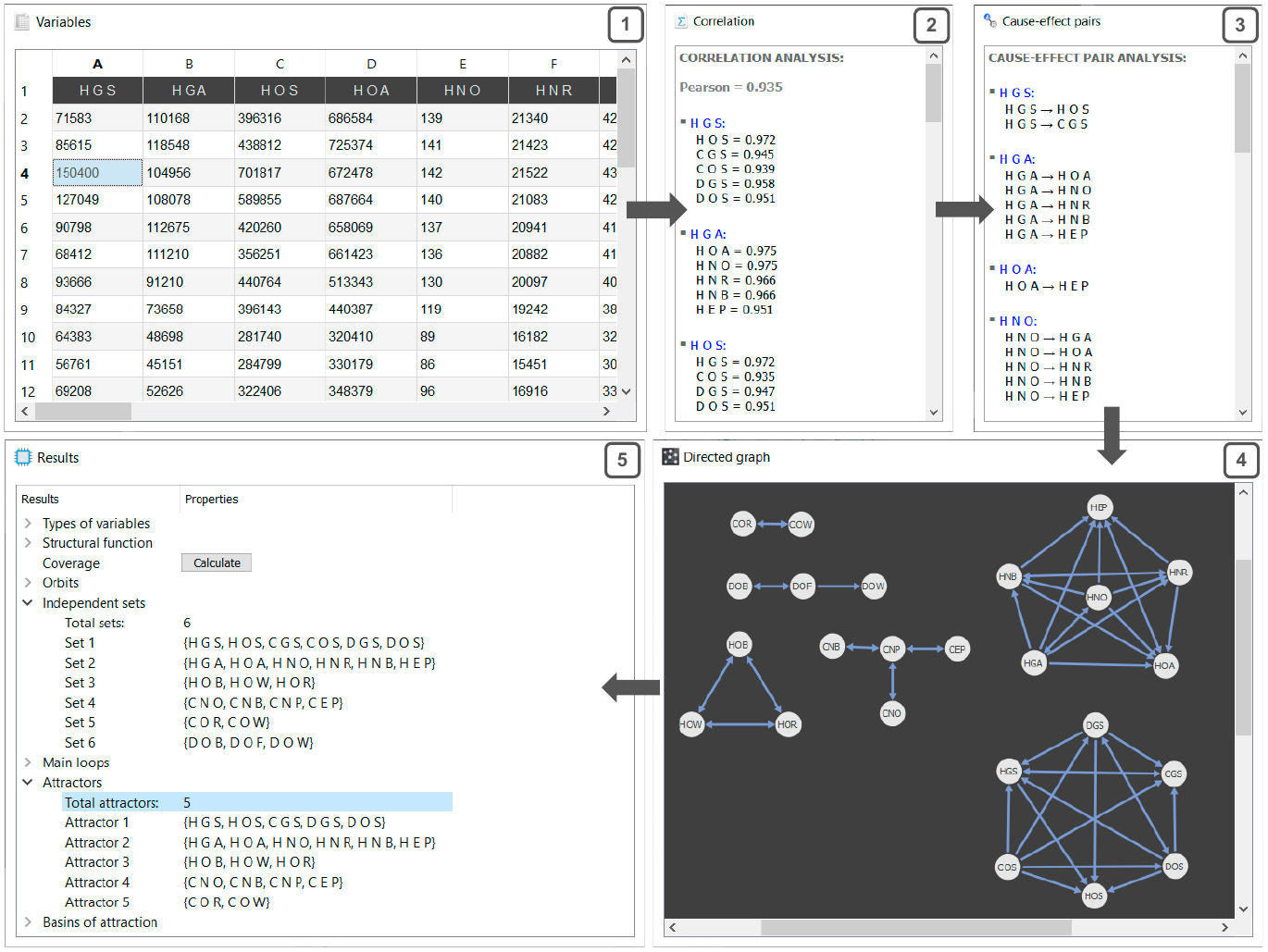}
	\caption{Work process in the Smarta software.}
	\label{fig:smarta}
\end{figure}

Next, once the above data was prepared, we used Smarta, which allows determining the trends of Benidorm’s tourism system and locating its different attractors over the study period. Thus, after entering the data from the surveys in the software (Fig. \ref{fig:smarta}, step 1), we proceeded to calculate Pearson’s correlation coefficients ($r$). To do this, we selected a correlation coefficient whose absolute value was greater than or equal to 0.935 (Fig. \ref{fig:smarta}, step 2).

We chose this correlation threshold because, on the one hand, a very high threshold allows identifying the system’s strongest relations; and on the other, thanks to this value, we can achieve an optimal balance, obtaining the greatest number of attractors including as many variables as possible, while sacrificing the least possible number of related cause-effect pairs. To use a simile, we eliminated the trees that stopped us from seeing the forest.

\begin{table}[t!]
	\centering
	\caption{Correlations obtained for variable pairs with $\left| r \right| \geqslant 0.935$.}
	\label{tab:2}
	{\scriptsize
		\begin{tabular}{lcllcllc}
			\toprule
			\textit{Correlated pair} & {[}$ r ${]}        &  & \textit{Correlated pair} & {[}$ r ${]}        &  & \textit{Correlated pair} & {[}$ r ${]}        \\
			\midrule
			HGS--HOS:&\textbf{0.972}&&HNR--HNO:&\textbf{0.983}&&COS--DGS:&\textbf{0.948}\\
			HGS--CGS:&\textbf{0.945}&&HNR--HNB:&\textbf{0.998}&&COS--DOS:&\textbf{0.971}\\
			HGS--COS:&\textbf{0.939}&&HNR--HEP:&\textbf{0.959}&&CNO--CNP:&\textbf{0.956}\\
			HGS--DGS:&\textbf{0.958}&&HNB--HGA:&\textbf{0.966}&&CNB--CNP:&\textbf{0.985}\\
			HGS--DOS:&\textbf{0.951}&&HNB--HOA:&\textbf{0.948}&&CNP--CNO:&\textbf{0.956}\\
			HGA--HOA:&\textbf{0.975}&&HNB--HNO:&\textbf{0.981}&&CNP--CNB:&\textbf{0.985}\\
			HGA--HNO:&\textbf{0.975}&&HNB--HNR:&\textbf{0.998}&&CNP--CEP:&\textbf{0.935}\\
			HGA--HNR:&\textbf{0.966}&&HNB--HEP:&\textbf{0.963}&&COR--COW:&\textbf{0.954}\\
			HGA--HNB:&\textbf{0.966}&&HOB--HOW:&\textbf{0.972}&&COW--COR:&\textbf{0.954}\\
			HGA--HEP:&\textbf{0.951}&&HOB--HOR:&\textbf{0.984}&&CEP--CNP:&\textbf{0.935}\\
			HOS--HGS:&\textbf{0.972}&&HOW--HOB:&\textbf{0.972}&&DGS--HGS:&\textbf{0.958}\\
			HOS--COS:&\textbf{0.935}&&HOW--HOR:&\textbf{0.959}&&DGS--HOS:&\textbf{0.947}\\
			HOS--DGS:&\textbf{0.947}&&HOR--HOB:&\textbf{0.984}&&DGS--CGS:&\textbf{0.936}\\
			HOS--DOS:&\textbf{0.951}&&HOR--HOW:&\textbf{0.959}&&DGS--COS:&\textbf{0.948}\\
			HOA--HGA:&\textbf{0.975}&&HEP--HGA:&\textbf{0.951}&&DGS--DOS:&\textbf{0.982}\\
			HOA--HNO:&\textbf{0.965}&&HEP--HOA:&\textbf{0.973}&&DOS--HGS:&\textbf{0.951}\\
			HOA--HNR:&\textbf{0.952}&&HEP--HNO:&\textbf{0.954}&&DOS--HOS:&\textbf{0.951}\\
			HOA--HNB:&\textbf{0.948}&&HEP--HNR:&\textbf{0.959}&&DOS--CGS:&\textbf{0.952}\\
			HOA--HEP:&\textbf{0.973}&&HEP--HNB:&\textbf{0.963}&&DOS--COS:&\textbf{0.971}\\
			HNO--HGA:&\textbf{0.975}&&CGS--HGS:&\textbf{0.945}&&DOS--DGS:&\textbf{0.982}\\
			HNO--HOA:&\textbf{0.965}&&CGS--COS:&\textbf{0.963}&&DNB--DNF:&\textbf{0.965}\\
			HNO--HNR:&\textbf{0.983}&&CGS--DGS:&\textbf{0.936}&&DNF--DNB:&\textbf{0.965}\\
			HNO--HNB:&\textbf{0.981}&&CGS--DOS:&\textbf{0.952}&&DOB--DOF:&\textbf{0.943}\\
			HNO--HEP:&\textbf{0.954}&&COS--HGS:&\textbf{0.939}&&DOF--DOB:&\textbf{0.943}\\
			HNR--HGA:&\textbf{0.966}&&COS--HOS:&\textbf{0.935}&&DOF--DOW:&\textbf{0.985}\\
			HNR--HOA:&\textbf{0.952}&&COS--CGS:&\textbf{0.963}&&DOW--DOF:&\textbf{0.985}\\
			\bottomrule
		\end{tabular}
	}
\end{table}

\begin{table}[t!]
	\centering
	\caption{Cause-effect pairs computed with $\left| r \right| \geqslant 0.935$.}
	\label{tab:3}
	{\scriptsize
		\begin{tabular}{lll}
			\toprule
			\multicolumn{3}{c}{\textit{Cause-effect pairs}}\\
			\midrule
			HGS $\rightarrow$ HOS & HNB $\rightarrow$ HNR & CNP $\rightarrow$ CNO \\
			HGS $\rightarrow$ CGS & HNB $\rightarrow$ HEP & CNP $\rightarrow$ CNB \\
			HGA $\rightarrow$ HOA & HOB $\rightarrow$ HOW & CNP $\rightarrow$ CEP \\
			HGA $\rightarrow$ HNO & HOB $\rightarrow$ HOR & COR $\rightarrow$ COW \\
			HGA $\rightarrow$ HNR & HOW $\rightarrow$ HOB & COW $\rightarrow$ COR \\
			HGA $\rightarrow$ HNB & HOW $\rightarrow$ HOR & CEP $\rightarrow$ CNP \\
			HGA $\rightarrow$ HEP & HOR $\rightarrow$ HOB & DGS $\rightarrow$ HGS \\
			HOA $\rightarrow$ HEP & HOR $\rightarrow$ HOW & DGS $\rightarrow$ HOS \\
			HNO $\rightarrow$ HGA & HEP $\rightarrow$ HOA & DGS $\rightarrow$ CGS \\
			HNO $\rightarrow$ HOA & CGS $\rightarrow$ HGS & DGS $\rightarrow$ DOS \\
			HNO $\rightarrow$ HNR & COS $\rightarrow$ HGS & DOS $\rightarrow$ HGS \\
			HNO $\rightarrow$ HNB & COS $\rightarrow$ HOS & DOS $\rightarrow$ HOS \\
			HNO $\rightarrow$ HEP & COS $\rightarrow$ CGS & DOS $\rightarrow$ CGS \\
			HNR $\rightarrow$ HOA & COS $\rightarrow$ DGS & DOS $\rightarrow$ DGS \\
			HNR $\rightarrow$ HNB & COS $\rightarrow$ DOS & DOB $\rightarrow$ DOF \\
			HNR $\rightarrow$ HEP & CNO $\rightarrow$ CNP & DOF $\rightarrow$ DOB \\
			HNB $\rightarrow$ HOA & CNB $\rightarrow$ CNP & DOF $\rightarrow$ DOW \\
			\bottomrule
		\end{tabular}
	}
\end{table}

Of the total 35 variables selected for our analysis, 26 exceeded the correlation threshold by one or more variables. Therefore, the 9 remaining variables---that did not exceed the correlation threshold---were left out of the analysis: HAS, CGA, COA, COP, CAS, DGA, DOA, DEP and DAS. Table \ref{tab:2} shows the results of the correlations obtained for the 26 variables that exceeded the threshold.

Next, we proceeded to calculate the cause-effect pairs (Fig. \ref{fig:smarta}, step 3). This time, we obtained a total of 51 cause-effect pairs. All calculated pairs are gathered in Table \ref{tab:3}.

\begin{figure}[t!]
	\centering
	\includegraphics[width=\columnwidth]{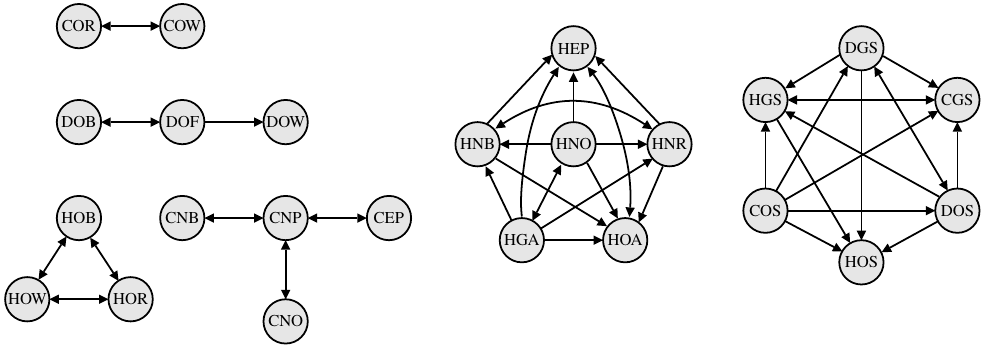}
	\caption{Directed graph of the databases of the Benidorm occupancy survey (period Jan. 2016 - Oct. 2018).}
	\label{fig:graph}
\end{figure}

Then, we represented the directed graph (Fig. \ref{fig:smarta}, step 4), symbolising each variable in the graph with a circle (the graph’s vertices), while each relation was illustrated with an arrow (the graph’s edges). Fig. \ref{fig:graph} represents the directed graph obtained for Benidorm’s occupancy survey databases. Upon analysis of the graph (Fig. \ref{fig:smarta}, step 5), we found a total of 6 independent sets. These sets (from $S_1$ to $S_6$) were as follow:

\begin{equation}
\begin{aligned}
{S_1} &= \left\{ {{\text{DGS}}{\text{, CGS}}{\text{, DOS}}{\text{, HOS}}{\text{, COS}}{\text{, HGS}}} \right\} \hfill \\
{S_2} &= \left\{ {{\text{HEP}}{\text{, HNR}}{\text{, HOA}}{\text{, HGA}}{\text{, HNB}}{\text{, HNO}}} \right\} \hfill \\
{S_3} &= \left\{ {{\text{HOB}}{\text{, HOR}}{\text{, HOW}}} \right\} \hfill \\
{S_4} &= \left\{ {{\text{CNB}}{\text{, CNP}}{\text{, CEP}}{\text{, CNO}}} \right\} \hfill \\
{S_5} &= \left\{ {{\text{DOB}}{\text{, DOF}}{\text{, DOW}}} \right\} \hfill \\
{S_6} &= \left\{ {{\text{COR}}{\text{, COW}}} \right\}
\end{aligned}
\end{equation}

Each set $S_i$ ($i=1,2,\ldots,6$) would generate its corresponding attractor $A_i$, where:

\begin{equation}
\begin{aligned}
{A_1} &= {S_1} - \left\{ {{\text{COS}}} \right\} = \left\{ {{\text{DGS}}{\text{, CGS}}{\text{, DOS}}{\text{, HOS}}{\text{, HGS}}} \right\} \hfill \\
{A_2} &= {S_2} = \left\{ {{\text{HEP}}{\text{, HNR}}{\text{, HOA}}{\text{, HGA}}{\text{, HNB}}{\text{, HNO}}} \right\} \hfill \\
{A_3} &= {S_3} = \left\{ {{\text{HOB}}{\text{, HOR}}{\text{, HOW}}} \right\} \hfill \\
{A_4} &= {S_4} = \left\{ {{\text{CNB}}{\text{, CNP}}{\text{, CEP}}{\text{, CNO}}} \right\} \hfill \\
{A_5} &= {S_5} = \left\{ {{\text{DOB}}{\text{, DOF}}{\text{, DOW}}} \right\} \hfill \\
{A_6} &= {S_6} = \left\{ {{\text{COR}}{\text{, COW}}} \right\}
\end{aligned}
\end{equation}

If we now study the obtained sets of attractors, we can deduce that attractor $A_1$ links the number of Spanish resident travellers to the number of overnight stays, for all 3 types of accommodation. Thus, an increase in the number of Spanish guests would positively influence the number of total overnight stays, and vice versa. Moreover, it is worth noting that the COS variable (overnight stays in Spanish resident campsites) disappears from the $A_1$ attractor even though it is present in $S_1$. This fact could reflect the fact that the share of overnight stays in campsites is a minority compared to that of hotels and holiday flats over the period Jan. 2016 - Oct. 2018 \cite{ine_campsite,femenia2021}. Specifically, during this period and in Benidorm city, hotels accounted for the highest share of overnight stays of the total (87.38\%), followed by holiday flats (9.52\%), and finally, campsites (3.10\%).

The $A_2$ attractor links travellers and foreign residents’ overnight stays to the hotel supply and employed staff. This set shows that the number of overseas guests in hotels (HGA) would be the cause of different variables related to the hotel offer, such as the number of open establishments (HNO), the number of rooms (HNR), the number of beds (HNB) and the personnel employed (HEP). On the other hand, the number of overnight stays of foreigners in the hotel sector (HOA) would directly result from the number of travellers and the supply variables commented above. This would make sense, since a greater number of travellers would cause an increase in the hotel offer and, in turn, the total number of nights that guests stay would also increase. In addition, interestingly, all the hotel sector’s supply variables are linked to overseas travellers (attractor $A_2$), but not to Spanish travellers (attractor $A_1$). This could be related to the fact that since 2016, tourism from abroad has contributed more to Benidorm’s hotel sector than national tourism \cite{ine_hotels, benidorm2018}. For example, in 2017, the percentage of overseas travellers reached 51.3\% of the total, while that of Spanish travellers was 48.7\%. Similarly, the percentage with respect to the total number of overnight stays was 57.3\% for guests from abroad, and 42.7\% for Spanish guests.

The $A_3$ attractor connects the 3 occupancy variables in hotels: bed occupancy rate (HOB), bed occupancy on weekends (HOW) and room occupancy (HOR). All these variables would form a bidirectional causal circle, such that any occupancy rate increase would positively affect the 2 remaining variables. This tells us that the 3 indicators would be equivalent to each other, and that they follow their own trend with respect to the rest of variable groups, since they form an independent attractor. In addition, from 2016 to 2019, occupancy rates per accommodation type were: 79.29\% for hotels (rooms), 82.63\% for campsites (plots) and 60.97\% for holiday apartments (flats) \cite{ine_hotels,ortuno2018}. This data was compared to the 3 attractors obtained for the occupancy levels ($A_3$, $A_5$ and $A_6$), one for each type of accommodation.

If we examine the $A_4$ attractor, we can see that it relates supply to the personnel employed on the campsites. In this case, the CNP variable (number of plots) would act as a bridge between the variables CNO (number of establishments), CNB (number of beds) and CEP (employed personnel). Thus, since all the attractor’s cause-effect pairs present bidirectional relationships, a positive or negative variation, for example, of the number of campsites available would generate a corresponding rise or drop in the number of total plots and beds, and consequently, a greater or lesser amount of personnel employed. On the other hand, it is worth noting that no variables are related to travellers or overnight stays in this attractor. One possible reason could be the fact that in the case of campsites, in 2017, the supply and personnel variables showed a more constant behaviour regarding annual variation, unlike travellers and overnight stay variables, whose annual variation was greater \cite{ine_campsite, leivestad2017}. This fact would cause one group of variables to be insufficiently correlated with another and not be able to surpass the established threshold of 0.935.

Finally, the last two attractors, $A_5$ and $A_6$, link holiday flats and campsites occupancy rate variables, respectively \cite{ine_campsite, ine_dwellings}. In the case of $A_5$, the variables DOB (bed occupancy rate) and DOF (flat occupancy rate) are found to causally feed back one another, so a higher rate of occupied flats would imply a greater use of beds, and vice versa. On the other hand, the DOF variable (flat occupancy rate) would be the cause of DOW (flat occupancy rate at weekends), which would make sense, since DOW would be part of the DOF. With regard to the attractor $A_6$, we can observe that the variables COR (occupancy rates by plots) and COW (plot occupancy levels at weekends) would be causally related: the occupancy increase of total plots includes an occupancy increase at weekends, and vice versa.

\subsection{The system’s tourism life cycle}
\label{subsec:tourism_life_cycle}

A distinct feature of Benidorm’s tourism system is that it has succeeded in adapting to the different phases that have unfolded since the end of the 80s. A major stage in the destination’s tourist indicator drop was the international 2007-2009 economic crisis, globally characterised by a substantial decline in the number of trips of traditionally international countries \cite{ivars2013}. This stage could be compared to the decline phase in Butler's model. Following this stage, Benidorm’s tourism variables have grown over the last few years. It has positioned itself in the Butler model’s rejuvenation stage, which can be justified by the fact that the destination learns from experience, sharing many features of complex adaptive systems \cite{levin2003}.

In this sense, the rejuvenation phase becomes clear upon examination of the most representative variables of each attractor found (from $A_1$ to $A_6$). The virtuous circles they present complement each other: entities cooperate through intergroups and intragroups, and the advantages of collaborating represent competitive advantages \cite{ritchie2003,gutierrez2018,crotti2017}.

\begin{figure}[t!]
	\centering
	\includegraphics[width=\columnwidth]{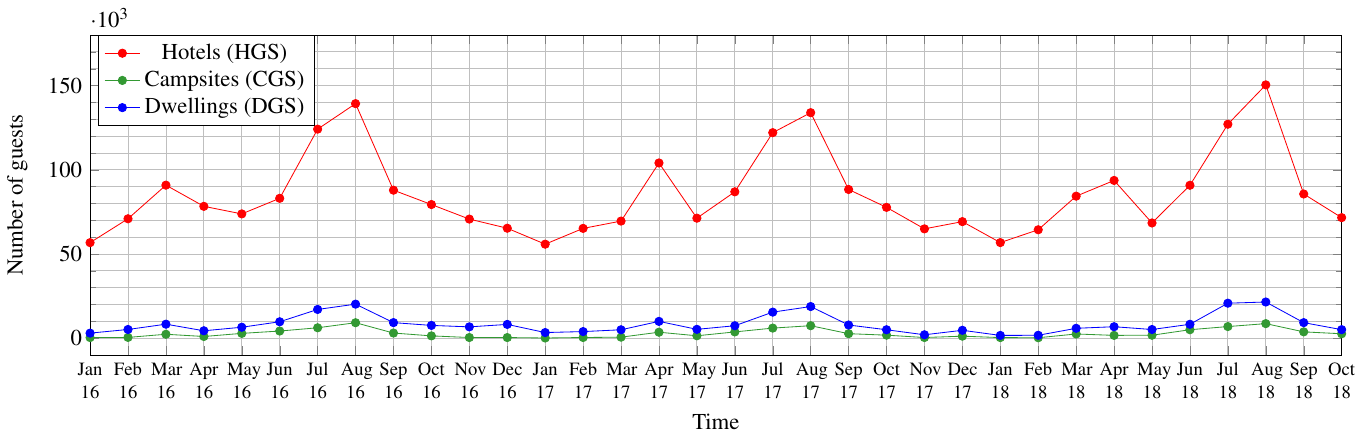}
	\caption{Representation of the significant variables of the $A_1$ attractor.}
	\label{fig:attractor1}
\end{figure}

\begin{figure}[t!]
	\centering
	\includegraphics[width=\columnwidth]{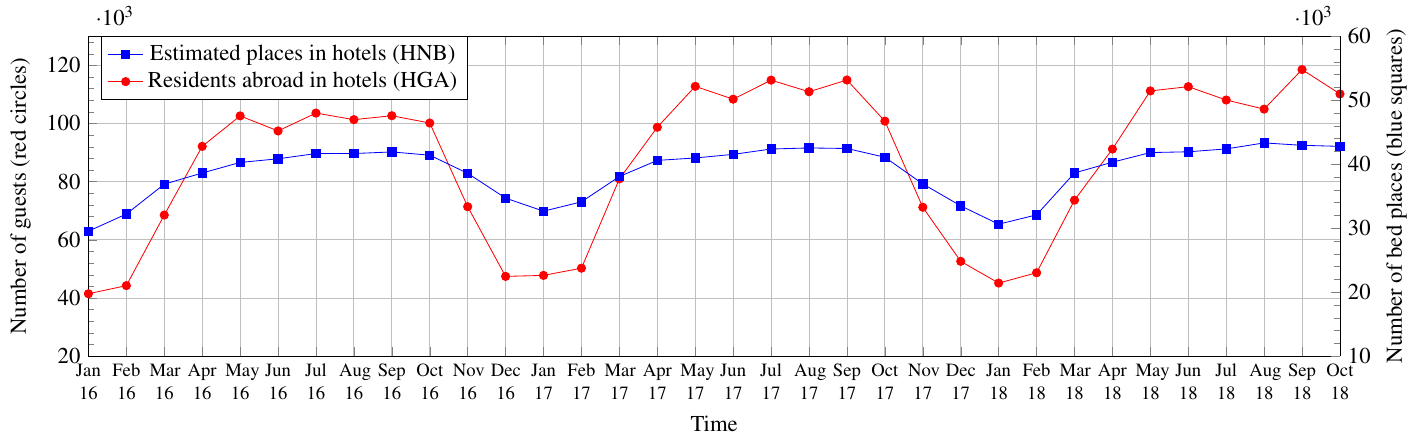}
	\caption{Representation of the significant variables of the $A_2$ attractor.}
	\label{fig:attractor2}
\end{figure}

In the case of the $A_1$ attractor, the significant variables we selected were those indicating the number of Spanish resident travellers for the three types of accommodation: hotels (HGS), campsites (CGS) and flats (DGS). Worthy of note, we discarded the variables of this attractor that referred to the number of overnight stays, since the latter would be directly related to the numbers of travellers variables. Therefore, when representing the variables HGS, CGS and DGS for the selected study period (Jan. 2016 - Oct. 2018), we would obtain the graph in Fig. \ref{fig:attractor1}. As expected, the maximum values of the number of travellers were reached during the high season months (March, April, June, July, August and September), and were always higher for hotels and smaller for campsites. This behaviour was perpetuated periodically during the rejuvenation phase, with a period $T = 12$ months. A slight upward trend was found. We can thus consider that it is the hotel sector that favours the rejuvenation of Benidorm’s tourism life cycle.

With regard to the $A_2$ attractor, we chose the variables that indicated the number of travellers from abroad (HGA) and the number of estimated beds (HNB), both referring to hotels. In Fig. \ref{fig:attractor2}, the left vertical axis measures the HGA variable (in red), while the right vertical axis measures the HNB variable (in blue). In the case of overseas travellers, the high season covers a more extended period than in the case of Spanish resident travellers. This stage begins in April, ends in October, and the behaviour repeats itself on an annual basis. In addition, we can appreciate the drop produced both in the number of travellers and in the number of beds available from November to March. Furthermore, the graph provides several other interesting insights:

\begin{itemize}
	\item The fact that the HGA variable scale is double that of the HNB would indicate that Benidorm’s hotel sector reaches high occupancy levels since there are many travellers for the number of available beds.
	\item Looking at the study period, the number of overseas travellers follows an upward trend. However, the number of beds did not increase in equal measure, probably due to the saturation of maximum hotel capacity during the high season.
	\item The absolute maximum in the number of overseas travellers (for the period Jan. 2016 - Oct. 2018) is reached in September 2018, with a total of 118,548 guests.
\end{itemize}

\begin{figure}[t!]
	\centering
	\includegraphics[width=\columnwidth]{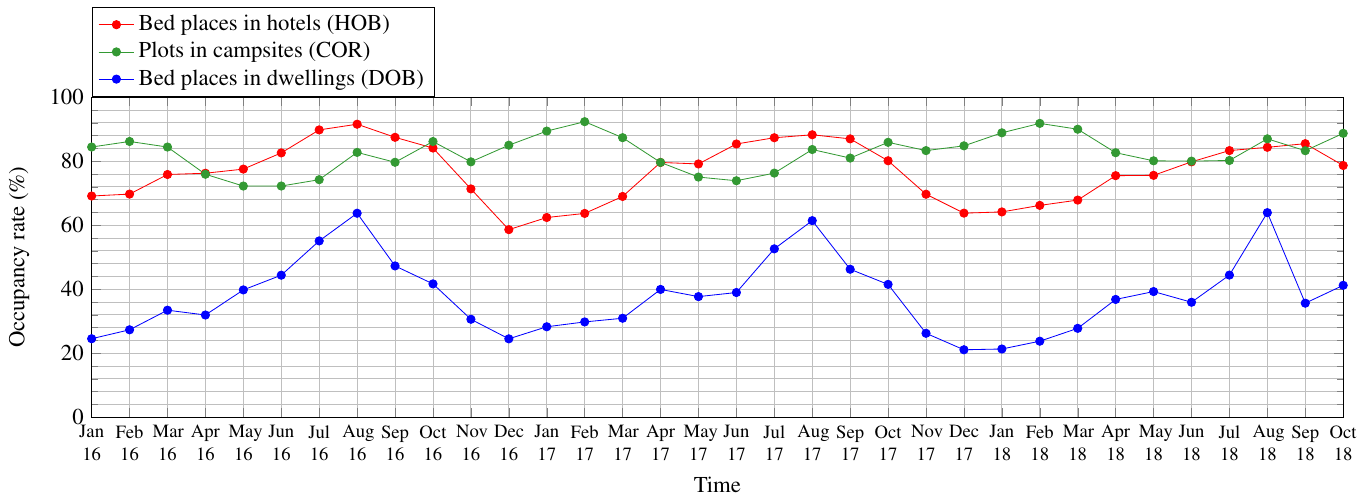}
	\caption{Representation of the significant variables of the $A_3$, $A_5$ and $A_6$ attractors.}
	\label{fig:attractor356}
\end{figure}

\begin{figure}[t!]
	\centering
	\includegraphics[width=\columnwidth]{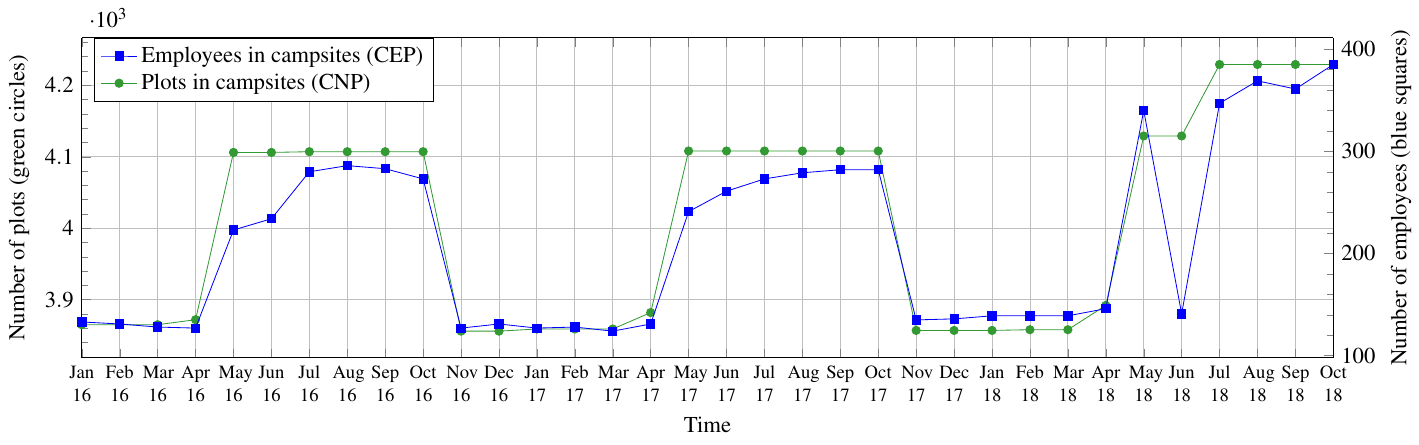}
	\caption{Representation of the significant variables of the $A_4$ attractor.}
	\label{fig:attractor4}
\end{figure}

Fig. \ref{fig:attractor356} shows the occupancy rate of the three types of accommodation. The 3 variables represented, HOB (bed occupancy of hotels), COR (campsites occupancy by plots) and DOB (flat occupancy by beds), would correspond to the most representative attractors $A_3$, $A_5$ and $A_6$, respectively. On the one hand, it is worth noting that similar trends are followed by both the hotel and flat sectors: occupancy rates increase notably during the months of May-October. However, in the case of hotels, higher percentages are obtained, as well as a lower annual variation, unlike the holiday apartment sector, that presents smaller percentages and more abrupt annual changes. On the other hand, we must bear in mind that the camping sector follows an opposite behaviour, that is, the months between August and April present high occupancy rates. It is also significant that campsites maintain more homogeneous occupancy rates throughout the year, surpassing even the hotel sector during the low season months (from November to April). This phenomenon is known as \textit{deseasonalization} and is mainly due to bookings from travellers from Europe (the highest percentage coming from the United Kingdom, followed by Holland, Belgium, Germany and France) during the winter months.

Finally, Fig. \ref{fig:attractor4} shows the $A_4$ attractor’s two most relevant variables, namely: the number of campsite plots available (CNP) and the amount of personnel employed for this type of accommodation (CEP). The graph’s left vertical axis represents the CNP variable (curve with green circles), while the right vertical axis represents the CEP variable (curve with blue squares). As we can observe, the number of plots rose from May to October (to meet high season demand), while from January to April, and from November to December, the number of plots dropped, due to reduced traveller influx. From January 2016 to March 2018, the number of plots reached maximum values close to 4,100 plots, and minimum values of approximately 3,800 plots. In contrast, as of May 2018, the number of available plots substantially increased. The fact that establishments increased their number of plots would respond to the camping sector’s current growth in the Benidorm area. Specifically, plots that include bungalows are usually the preferred option of families staying on campsites, so this type of accommodation has been greatly promoted, in addition to other types of plots, such as plots for tents and caravans. Finally, the graph describing the number of employees follows that of the number of plots adaptively, since the establishments adapt the amount of employed personnel to their customers’ needs. Worthy of note, although the number of available plots changes abruptly according to the high or low season, the number of employees undergoes a more gradual evolution. This reflects the fact that establishments do not hire a maximum number of employees at the beginning of the season: the number of employees changes according to each campsite’s current situation.

\section{Conclusions}
\label{sec:conclusions}

The attractors obtained for the selected study period constitute an indicator of the rejuvenation phase of Benidorm’s tourism life cycle. A possible way of explaining this behaviour is that the system learns from experience and acts as a complex adaptive system. This conclusion is supported by the fact that, over the period considered, six attractors were obtained; they were determined by virtuous cycles that analyze different parts of the system and complement each other. There is a degree of cooperation between the entities through intergroups and intragroups, as shown by the cycles obtained. We can thus consider that the advantages of collaboration represent competitive advantages and the destination has achieved systemic competitiveness. The obtained periodicity, combined with an upward trend in the graphs, shows that these variables followed similar behaviours over all the years. They also reflect that the calculated attractors maintain the stability of the relationships, allowing the destination to develop positively.

Regarding the limitations of the present study, we can highlight that, although the technique presented helps us to detect possible causal relationships in tourism systems, there could be unknown factors that the technique does not contemplate, which could affect the tourism variables. Another limitation would be that the study did not consider rural lodgings or hostels, as they are a minority in Benidorm. A final limitation would be that, at present, Smarta can only work with quantitative variables, but not categorical ones (such as, for example, the type of accommodation preferred by a guest).

In terms of future research, we would like to prepare an article analyzing tourism variables for the period 2019-2021, with the aim of comparing trends, attractors and the tourism life cycle of Benidorm, during the pre-COVID and post-COVID era. Complementarily, it would also be interesting to study how Brexit has affected this tourist destination, so dependent on British tourism. Finally, another line of work could be how to optimize the network associated with Benidorm's tourism sector.


\bibliographystyle{elsarticle-num} 
\bibliography{bibliography}

%
%
%
\end{document}